\documentclass[twocolumn,amssymb, nobibnotes, aps, prd,amsmath,mathtools, nofootinbib,pgfmath,pgffor,superscriptaddress]{revtex4-1}
\usepackage{graphicx}% Include figure files

\usepackage{caption}
\usepackage{subcaption}

\usepackage{hyperref}
\hypersetup{colorlinks,citecolor=blue}

\makeatletter
\def\qrr@split@result#1 #2\@qrr@split@result{\edef\erfInput{#1}\edef\erfResult{#2}}
\newcommand*{\gnuplotErf}[2][\jobname.eval]{%
    \immediate\write18{gnuplot -e "set print '#1'; print #2, erf(#2);"}%
    \everyeof{\noexpand}
    \edef\qrr@temp{\input{#1} }%
    \expandafter\qrr@split@result\qrr@temp\@qrr@split@result
}
\makeatother

\usepackage{caption,color}
\usepackage{subcaption}
\usepackage{multirow}

\usepackage{hyperref}
\hypersetup{colorlinks,citecolor=blue}

\begin{document}

\title{Comment on:\\ ``Low significance of evidence for black hole echoes in gravitational wave data''}

\author{Jahed Abedi}
\email{jahed$_$abedi@physics.sharif.ir}
\affiliation{Department of Physics, Sharif University of Technology, P.O. Box 11155-9161, Tehran, Iran}
\affiliation{School of Particles and Accelerators, Institute for Research in Fundamental Sciences (IPM), P.O. Box 19395-5531, Tehran, Iran}
%\affiliation{Perimeter Institute for Theoretical Physics, 31 Caroline St. N., Waterloo, ON, N2L 2Y5, Canada}

\author{Hannah Dykaar}
\affiliation{Department of Physics, McGill University, 3600 rue University, Montreal, QC, H3A 2T8, Canada}
%\affiliation{Department of Physics and Astronomy, University of Waterloo, Waterloo, ON, N2L 3G1, Canada}

\author{Niayesh Afshordi}
\email{nafshordi@pitp.ca}
\affiliation{Perimeter Institute for Theoretical Physics, 31 Caroline St. N., Waterloo, ON, N2L 2Y5, Canada}
\affiliation{Department of Physics and Astronomy, University of Waterloo, Waterloo, ON, N2L 3G1, Canada}

%\date{June 16, 2015}
%\date{\today}
%--------------------------------------------------------------------------
\begin{abstract}
In a recent publication \cite{Abedi:2016hgu}, we demonstrated that the events in the first observing run of the Advanced LIGO gravitational wave observatory (aLIGO O1) showed tentative evidence for repeating ``echoes from the abyss'' caused by Planck-scale structure near black hole horizons. By considering a phenomenological echo model, we showed that the pure noise hypothesis is disfavored with a p-value of 1\%, i.e. higher amplitude for echoes than those in aLIGO O1 events are only recovered in 1\% of random noise realizations. A recent preprint by Westerweck, et al. \cite{Westerweck:2017hus} provides a careful re-evaluation of our analysis which claims ``a reduced statistical significance ... entirely consistent with noise''. It is a mystery to us why the authors make such a statement, while they also find a p-value of $2 \pm 1\%$ (given the Poisson error in their estimate) for the same model and dataset. This is p-erfectly consistent with our results, which would be commonly considered as disfavoring the null hypothesis, or ``moderate to significant'' evidence for ``echoes''. Westerweck, et al. \cite{Westerweck:2017hus} also point to diversity of the observed echo properties as evidence for statistical fluke, but such a diversity is neither unique nor surprising for complex physical phenomena in nature. 
\end{abstract}

\maketitle
%\tableofcontents

The recent arXiv preprint entitled ``Low significance of evidence for black hole echoes in gravitational wave data'' by Westerweck et al. \cite{Westerweck:2017hus} provides a very thorough and careful reevaluation of the claims in our article \cite{Abedi:2016hgu}, which presented tentative evidence for repeating echoes in the gravitational wave observations of binary black hole mergers, with period of
\begin{eqnarray}
&\Delta t_{echo} \simeq \frac{4 G M_{\rm BH}}{c^3}\left(1+\frac{1}{\sqrt{1-a^2}}\right) \times \ln\left(M_{\rm BH} \over M_{\rm planck}\right)& \nonumber\\ &\simeq 0.126~ {\rm sec} \left(M_{\rm BH} \over 67 ~ M_\odot \right) \left(1+\frac{1}{\sqrt{1-a^2}}\right),& \label{delay}
\end{eqnarray} 
due to reflections off putative Planck-scale  structure near horizon of the final black hole with mass $M_{\rm BH}$ and dimensionless spin $a$.

We have been aware of, and keenly following the analysis of Westerweck et al. (most notably, during the workshop \href{https://www.perimeterinstitute.ca/conferences/quantum-black-holes-sky}{``Quantum Black Holes in the Sky?''} held at Perimeter Institute in November 2017). Indeed, our own analysis in  \cite{Abedi:2016hgu} has already been significantly improved through these interactions (compared to its original version) \cite{Abedi:2017isz,Ashton:2016xff}. As a result, we have a very high opinion of the analysis by Westerweck et al., but are perplexed by why their Abstract/Conclusions misrepresent their findings. 

The strongest and ultimate statement in the Abstract is: ``The reduced significance is entirely consistent with noise, and so we conclude that the analysis of Abedi et al. does not provide any observational evidence for the existence of Planck-scale structure at black hole horizons.''

However, their Table I shows the p-value for the noise hypothesis, using the model and data used in Abedi et al. as 0.020 (as opposed to 0.011 in Abedi et al.  \cite{Abedi:2016hgu}). By most accounts (e.g., \href{https://en.wikipedia.org/wiki/P-value#Usage}{https://en.wikipedia.org/wiki/P-value\#Usage}), this would be considered a rejection of noise hypothesis (or ``moderate to strong'' Bayesian evidence for echoes \cite{goodman2001p}), i.e. the {\it opposite} of what their Abstract says. 

In fact, despite the laundry list of shortcomings of  \cite{Abedi:2016hgu} discussed by Westerweck et al., there is NO evidence that their improved analysis has reduced the significance, as their p-value of 0.020 is based on 5 (out of 250) background peaks that exceed the observed signal. However, the Poisson noise on 5 discrete events is $\sqrt{5}$, implying that their estimate of p-value is $0.020 \pm 0.009$, entirely consistent with $0.011$ of Abedi et al.   

There are certainly peculiarities about why certain events contribute to the echo signal, or the best-fit parameters, which are important to study and understand. However, as we elaborate below, they can only be evaluated in the context of a (yet non-existent) theoretical prior for echoes, and do not affect the statistical significance. Therefore, in their Abstract/Conclusions, Westerweck et al. do not separate objective statements about statistical significance, from subjective ones about their priors: If your coin comes up with 49 heads and 1 tail after 50 throws, then no one would say itÕs ``entirely consistent with noise'' for a fair coin.     

Below are more detailed comments that elaborate these points:
\begin{enumerate}
\item 	The only new dataset used by Westerweck et al. that was not available to us (at the time of our analysis) is GW 170104.  As such, we do not provide the range of $\Delta t_{echo}$ expected for this event (which would depend on the final detector frame mass and spin), and Westerweck et al. do not state the range that they used for their analysis either.  It should be noted that, if similar to previous events the 1-sigma range for $\Delta t_{echo}$ is used; there is 32\% probability that it misses the correct $\Delta t_{echo}$, which would lead to a reduced significance (by diluting the $\sum {\rm SNR}^2$ signal from previous events).
\item 	An important finding by Westerweck et al. is that the least significant LIGO event LVT151012 has the most contribution to (what we call) Òtentative evidenceÓ for echoes (even though the most significant LIGO BH event, GW150914 further improves the p-value by a factor of 3). While this is peculiar, it would only disfavor echoes if one believes the significance of echoes should scale with the significance of main event. However, there is simply no justification for this. For example, \cite{Wang:2018gin} shows that the SNR$^2$ for echoes can change by {\it 3 orders of magnitude} if the frequency of initial condition for echoes change by only $\pm 20\%$. As such, the relative echo signal can significantly vary over different BH binary events, as they have different component spins and mass ratios.  
\item	Westerweck et al. suggest that, since $\gamma \sim 1$ (with $\gamma$ being the ratio of subsequent echoes) is often preferred when fitting to pure noise, the fact that our best-fit $\gamma$ for echo signal is 0.9 is alarming. However, as it has been shown by e.g., \cite{Correia:2018apm,Wang:2018gin}, the echoes are expected to decay as power-laws (or inverse polynomials; with effective $\gamma \sim 1$), and thus finding $\gamma \sim 1$ cannot distinguish between the noise and echo hypotheses. 
\item	It would be very useful if Westerweck et al. could provide an analog of their Fig. 3, for $A_{\rm recovered}$ vs $A_{inj}$. The reason is that the procedure of maximizing SNR used by us and Westerweck et al. tends to give an overestimate of the best-fit amplitude. This would explain why LVT151012 has a bigger recovered amplitude than GW150914, as searching a larger range of parameter space (for more uncertain $\Delta t_{echo}$) tends to find bigger SNR peaks and thus bigger amplitudes. There is still a noticeable bump on top of the higher background (hence the low p-value), but this is why the background for SNR$_{\rm max}$ is higher for a noisier event.  
\item	The discussion around: Òpotential contamination of the background samples by existing echo signals, ÉÓ amounts to contradiction in terms! Entire point of computing a p-value is to evaluate the noise hypothesis, i.e. how often pure noise in data can mimic echo signal, while one doesnÕt exist. If you assume there are already echoes in the data, then you have already rejected the noise hypothesis!
\item Westerweck et al. state: ``we find a p-value for the null noise only hypothesis of 0.02, higher than that reported using the restricted background of 0.011 in [20].'', except that they fail to report that Poisson error of 0.009 in their p-value estimation, which makes the two values consistent. This also suggests that our Òrestricted backgroundÓ, critiqued by Westerweck et al., may have been sufficient for  p-value estimation. 
\end{enumerate}

To conclude, while Westerweck et al.  \cite{Westerweck:2017hus}  provide a much more thorough statistical analysis in searching for echoes in LIGO data, they do reproduce our results in \cite{Abedi:2016hgu}. Their real criticism stems from a simplistic choice of prior for the echo signal, which in our view is unjustified. Of course, p-values of 1-2\% do not amount to ``detections'' for good reason. Fortunately, analysis of new LIGO events by independent methodologies and/or groups are improving this situation \cite{Conklin:2017lwb,Abedi:2018}. 

%--------------------------------------------------------------------------------------------
%--------------------------------------------------------------------------------------------
\bibliography{comment}

\providecommand{\noopsort}[1]{}\providecommand{\singleletter}[1]{#1}%
\begin{thebibliography}{9}%
\makeatletter
\providecommand \@ifxundefined [1]{%
 \@ifx{#1\undefined}
}%
\providecommand \@ifnum [1]{%
 \ifnum #1\expandafter \@firstoftwo
 \else \expandafter \@secondoftwo
 \fi
}%
\providecommand \@ifx [1]{%
 \ifx #1\expandafter \@firstoftwo
 \else \expandafter \@secondoftwo
 \fi
}%
\providecommand \natexlab [1]{#1}%
\providecommand \enquote  [1]{``#1''}%
\providecommand \bibnamefont  [1]{#1}%
\providecommand \bibfnamefont [1]{#1}%
\providecommand \citenamefont [1]{#1}%
\providecommand \href@noop [0]{\@secondoftwo}%
\providecommand \href [0]{\begingroup \@sanitize@url \@href}%
\providecommand \@href[1]{\@@startlink{#1}\@@href}%
\providecommand \@@href[1]{\endgroup#1\@@endlink}%
\providecommand \@sanitize@url [0]{\catcode `\\12\catcode `\$12\catcode
  `\&12\catcode `\#12\catcode `\^12\catcode `\_12\catcode `\%12\relax}%
\providecommand \@@startlink[1]{}%
\providecommand \@@endlink[0]{}%
\providecommand \url  [0]{\begingroup\@sanitize@url \@url }%
\providecommand \@url [1]{\endgroup\@href {#1}{\urlprefix }}%
\providecommand \urlprefix  [0]{URL }%
\providecommand \Eprint [0]{\href }%
\providecommand \doibase [0]{http://dx.doi.org/}%
\providecommand \selectlanguage [0]{\@gobble}%
\providecommand \bibinfo  [0]{\@secondoftwo}%
\providecommand \bibfield  [0]{\@secondoftwo}%
\providecommand \translation [1]{[#1]}%
\providecommand \BibitemOpen [0]{}%
\providecommand \bibitemStop [0]{}%
\providecommand \bibitemNoStop [0]{.\EOS\space}%
\providecommand \EOS [0]{\spacefactor3000\relax}%
\providecommand \BibitemShut  [1]{\csname bibitem#1\endcsname}%
\let\auto@bib@innerbib\@empty
%</preamble>
\bibitem [{\citenamefont {Abedi}\ \emph
  {et~al.}(2017{\natexlab{a}})\citenamefont {Abedi}, \citenamefont {Dykaar},\
  and\ \citenamefont {Afshordi}}]{Abedi:2016hgu}%
  \BibitemOpen
  \bibfield  {author} {\bibinfo {author} {\bibfnamefont {J.}~\bibnamefont
  {Abedi}}, \bibinfo {author} {\bibfnamefont {H.}~\bibnamefont {Dykaar}}, \
  and\ \bibinfo {author} {\bibfnamefont {N.}~\bibnamefont {Afshordi}},\ }\href
  {\doibase 10.1103/PhysRevD.96.082004} {\bibfield  {journal} {\bibinfo
  {journal} {Phys. Rev.}\ }\textbf {\bibinfo {volume} {D96}},\ \bibinfo {pages}
  {082004} (\bibinfo {year} {2017}{\natexlab{a}})},\ \Eprint
  {http://arxiv.org/abs/1612.00266} {arXiv:1612.00266 [gr-qc]} \BibitemShut
  {NoStop}%
%%CITATION = ARXIV:1612.00266;%%
\bibitem [{\citenamefont {Westerweck}\ \emph {et~al.}(2017)\citenamefont
  {Westerweck}, \citenamefont {Nielsen}, \citenamefont {Fischer-Birnholtz},
  \citenamefont {Cabero}, \citenamefont {Capano}, \citenamefont {Dent},
  \citenamefont {Krishnan}, \citenamefont {Meadors},\ and\ \citenamefont
  {Nitz}}]{Westerweck:2017hus}%
  \BibitemOpen
  \bibfield  {author} {\bibinfo {author} {\bibfnamefont {J.}~\bibnamefont
  {Westerweck}}, \bibinfo {author} {\bibfnamefont {A.}~\bibnamefont {Nielsen}},
  \bibinfo {author} {\bibfnamefont {O.}~\bibnamefont {Fischer-Birnholtz}},
  \bibinfo {author} {\bibfnamefont {M.}~\bibnamefont {Cabero}}, \bibinfo
  {author} {\bibfnamefont {C.}~\bibnamefont {Capano}}, \bibinfo {author}
  {\bibfnamefont {T.}~\bibnamefont {Dent}}, \bibinfo {author} {\bibfnamefont
  {B.}~\bibnamefont {Krishnan}}, \bibinfo {author} {\bibfnamefont
  {G.}~\bibnamefont {Meadors}}, \ and\ \bibinfo {author} {\bibfnamefont
  {A.~H.}\ \bibnamefont {Nitz}},\ }\href@noop {} {\  (\bibinfo {year}
  {2017})},\ \Eprint {http://arxiv.org/abs/1712.09966} {arXiv:1712.09966
  [gr-qc]} \BibitemShut {NoStop}%
%%CITATION = ARXIV:1712.09966;%%
\bibitem [{\citenamefont {Abedi}\ \emph
  {et~al.}(2017{\natexlab{b}})\citenamefont {Abedi}, \citenamefont {Dykaar},\
  and\ \citenamefont {Afshordi}}]{Abedi:2017isz}%
  \BibitemOpen
  \bibfield  {author} {\bibinfo {author} {\bibfnamefont {J.}~\bibnamefont
  {Abedi}}, \bibinfo {author} {\bibfnamefont {H.}~\bibnamefont {Dykaar}}, \
  and\ \bibinfo {author} {\bibfnamefont {N.}~\bibnamefont {Afshordi}},\
  }\href@noop {} {\  (\bibinfo {year} {2017}{\natexlab{b}})},\ \Eprint
  {http://arxiv.org/abs/1701.03485} {arXiv:1701.03485 [gr-qc]} \BibitemShut
  {NoStop}%
%%CITATION = ARXIV:1701.03485;%%
\bibitem [{\citenamefont {Ashton}\ \emph {et~al.}(2016)\citenamefont {Ashton},
  \citenamefont {Birnholtz}, \citenamefont {Cabero}, \citenamefont {Capano},
  \citenamefont {Dent}, \citenamefont {Krishnan}, \citenamefont {Meadors},
  \citenamefont {Nielsen}, \citenamefont {Nitz},\ and\ \citenamefont
  {Westerweck}}]{Ashton:2016xff}%
  \BibitemOpen
  \bibfield  {author} {\bibinfo {author} {\bibfnamefont {G.}~\bibnamefont
  {Ashton}}, \bibinfo {author} {\bibfnamefont {O.}~\bibnamefont {Birnholtz}},
  \bibinfo {author} {\bibfnamefont {M.}~\bibnamefont {Cabero}}, \bibinfo
  {author} {\bibfnamefont {C.}~\bibnamefont {Capano}}, \bibinfo {author}
  {\bibfnamefont {T.}~\bibnamefont {Dent}}, \bibinfo {author} {\bibfnamefont
  {B.}~\bibnamefont {Krishnan}}, \bibinfo {author} {\bibfnamefont {G.~D.}\
  \bibnamefont {Meadors}}, \bibinfo {author} {\bibfnamefont {A.~B.}\
  \bibnamefont {Nielsen}}, \bibinfo {author} {\bibfnamefont {A.}~\bibnamefont
  {Nitz}}, \ and\ \bibinfo {author} {\bibfnamefont {J.}~\bibnamefont
  {Westerweck}},\ }\href@noop {} {\  (\bibinfo {year} {2016})},\ \Eprint
  {http://arxiv.org/abs/1612.05625} {arXiv:1612.05625 [gr-qc]} \BibitemShut
  {NoStop}%
%%CITATION = ARXIV:1612.05625;%%
\bibitem [{\citenamefont {Goodman}(2001)}]{goodman2001p}%
  \BibitemOpen
  \bibfield  {author} {\bibinfo {author} {\bibfnamefont {S.~N.}\ \bibnamefont
  {Goodman}},\ }\href@noop {} {\bibfield  {journal} {\bibinfo  {journal}
  {Epidemiology}\ }\textbf {\bibinfo {volume} {12}},\ \bibinfo {pages} {295}
  (\bibinfo {year} {2001})}\BibitemShut {NoStop}%
\bibitem [{\citenamefont {Wang}\ and\ \citenamefont
  {Afshordi}(2018)}]{Wang:2018gin}%
  \BibitemOpen
  \bibfield  {author} {\bibinfo {author} {\bibfnamefont {Q.}~\bibnamefont
  {Wang}}\ and\ \bibinfo {author} {\bibfnamefont {N.}~\bibnamefont
  {Afshordi}},\ }\href@noop {} {\  (\bibinfo {year} {2018})},\ \Eprint
  {http://arxiv.org/abs/1803.02845} {arXiv:1803.02845 [gr-qc]} \BibitemShut
  {NoStop}%
%%CITATION = ARXIV:1803.02845;%%
\bibitem [{\citenamefont {Correia}\ and\ \citenamefont
  {Cardoso}(2018)}]{Correia:2018apm}%
  \BibitemOpen
  \bibfield  {author} {\bibinfo {author} {\bibfnamefont {M.~R.}\ \bibnamefont
  {Correia}}\ and\ \bibinfo {author} {\bibfnamefont {V.}~\bibnamefont
  {Cardoso}},\ }\href@noop {} {\  (\bibinfo {year} {2018})},\ \Eprint
  {http://arxiv.org/abs/1802.07735} {arXiv:1802.07735 [gr-qc]} \BibitemShut
  {NoStop}%
%%CITATION = ARXIV:1802.07735;%%
\bibitem [{\citenamefont {Conklin}\ \emph {et~al.}(2017)\citenamefont
  {Conklin}, \citenamefont {Holdom},\ and\ \citenamefont
  {Ren}}]{Conklin:2017lwb}%
  \BibitemOpen
  \bibfield  {author} {\bibinfo {author} {\bibfnamefont {R.~S.}\ \bibnamefont
  {Conklin}}, \bibinfo {author} {\bibfnamefont {B.}~\bibnamefont {Holdom}}, \
  and\ \bibinfo {author} {\bibfnamefont {J.}~\bibnamefont {Ren}},\ }\href@noop
  {} {\  (\bibinfo {year} {2017})},\ \Eprint {http://arxiv.org/abs/1712.06517}
  {arXiv:1712.06517 [gr-qc]} \BibitemShut {NoStop}%
%%CITATION = ARXIV:1712.06517;%%
\bibitem [{\citenamefont {Abedi}\ and\ \citenamefont
  {Afshordi}(2018)}]{Abedi:2018}%
  \BibitemOpen
  \bibfield  {author} {\bibinfo {author} {\bibfnamefont {J.}~\bibnamefont
  {Abedi}}\ and\ \bibinfo {author} {\bibfnamefont {N.}~\bibnamefont
  {Afshordi}},\ }\href@noop {} {\bibfield  {journal} {\bibinfo  {journal} {to
  appear}\ } (\bibinfo {year} {2018})}\BibitemShut {NoStop}%
\end{thebibliography}%
%\begin{thebibliography}{99}

%\end{thebibliography}

\end{document}